\documentclass[letterpaper, 10 pt, conference]{ieeeconf}  

\IEEEoverridecommandlockouts                              
\usepackage{graphics} 
\usepackage{epsfig} 
\usepackage{mathptmx} 
\usepackage{times} 
\usepackage{amsmath} 
\usepackage{authblk}
\usepackage{cite}
\usepackage{xcolor}
\usepackage[english, polish]{babel}
\title{\LARGE \bf
Synchronisation and calibration of the 24-modules J-PET prototype with 300~mm axial field of view
}

\author{
P.~Moskal$^{1}$, 
T.~Bednarski$^1$, 
Sz.~Nied\'zwiecki$^1$,
M.~Silarski$^1$,
E.~Czerwi\'nski$^1$,
T.~Kozik$^1$, 
J.~Chhokar$^1$,
M.~Ba{\l}a$^1$,\\
C.~Curceanu$^{2}$,
R.~Del~Grande$^2$, 
M.~Dadgar$^1$,
K.~Dulski$^1$, 
A.~Gajos$^1$, 
M.~Gorgol$^3$, 
N.~Gupta-Sharma$^{1}$,
B.~C.~Hiesmayr$^4$,\\
B.~Jasi{\'n}ska$^3$, 
K.~Kacprzak$^1$, 
\L.~Kap{\l}on$^1$,
H.~Karimi$^1$,
D.~Kisielewska$^1$,
K.~Klimaszewski$^5$,
G.~Korcyl$^1$, 
P.~Kowalski$^5$,  \\ 
N.~Krawczyk$^1$, 
W.~Krzemie\'n$^6$,  
E.~Kubicz$^1$, 
M.~Mohammed$^{1,7}$,  
M.~Pa{\l}ka$^1$,
M.~Pawlik-Nied{\'z}wiecka$^1$,
L.~Raczy\'nski$^5$, \\
J.~Raj$^1$, 
S.~Sharma$^1$, 
Shivani$^1$,
R.~Y.~Shopa$^5$,  
M.~Skurzok$^1$,
E.~St\c{e}pie{\'n}$^1$,
W.~Wi\'slicki$^5$, 
B.~Zgardzi{\'n}ska$^3$

\thanks{$^1$ Faculty of Physics, Astronomy and Applied Computer Science, Jagiellonian University, 30-348 Cracow }
\thanks{$^2$ INFN, Laboratori Nazionali di Frascati, 00044 Frascati, Italy}
\thanks{$^3$ Institute of Physics, Maria Curie-Sk\l odowska University, 20-031 Lublin, Poland}
\thanks{$^4$ Faculty of Physics, University of Vienna, 1090 Vienna, Austria}
\thanks{$^5$ Department of Complex Systems, National Centre for Nuclear Research, 05-400 Otwock-\'Swierk, Poland}
\thanks{$^6$ High Energy Physics Division, National Centre for Nuclear Research, 05-400 Otwock-\'Swierk, Poland}
\thanks{$^7$ Department of Physics, College of Education for Pure Sciences, University of Mosul, Mosul, Iraq
}}

\begin{document}
\bstctlcite{IEEEexample:BSTcontrol}
\maketitle
\thispagestyle{empty}
\pagestyle{empty}

\begin{abstract}
Research conducted in the framework of the J-PET project aims to develop a cost-effective total-body positron emission tomography scanner. As a first step on the way to
construct a full-scale J-PET tomograph from long strips of plastic scintillators, a 24-strip prototype was built and tested. The prototype consists of detection modules arranged axially forming a cylindrical diagnostic chamber with the inner diameter of 360~mm and the axial field-of-view of 300~mm. Promising perspectives for a low-cost construction of a total-body PET scanner are opened due to an axial arrangement of strips of plastic scintillators, wchich have a small light attenuation, superior timing properties, and the possibility of cost-effective increase of the axial field-of-view. The presented prototype comprises dedicated solely digital front-end electronic circuits and a triggerless data acquisition system which required development of new calibration methods including time, thresholds and gain synchronization. The system and elaborated calibration methods including first results of the 24-module J-PET prototype are presented and discussed. The achieved coincidence resolving time equals to CRT~=~490~$\pm$~9~ps. This value can be translated to the position reconstruction accuracy $\sigma(\Delta l) =$~18~mm which is fairly position-independent.

\end{abstract}

\section{Introduction}

Positron emission tomography (PET) is a well established molecular imaging technique\cite{JONES2017}.
It involves administration of  radiolabeled molecules containing elements emitting positrons to patients. Photons created due to the positron-electron annihilation are measured in order to localize and quantify the radiotracer~\cite{CHERRY2017A}. The principles underlying PET allow to study many biological processes e.g. metabolism (brain and cancer activity), hypoxia, apoptosis, proliferation (cancer), angiogenesis and inflammation (atherosclerotic plaque)\cite{JONES2017}. PET has been used  extensively for research and clinical applications, particularly concerning imaging of brain function in neurodegenerative diseases, diagnosis and treatment of cancer (theranostic) or monitoring of radio- and pharmacotherapy progress. By choosing different markers, one can select different metabolic processes that are observed during scanning. All  modern scanners, currently available on the market, detect $\gamma$-photons by usage of inorganic crystal scintillators~\cite{Karp2008a, SLOMKA2016, Van2016}. Scintillators in the form of crystal (eg. LSO or LYSO) are expensive but have undoubted advantages such as large density and high atomic number, and therefore a large cross-section for the interaction with annihilation photons through the photoelectric effect, in addition to a good energy resolution.
\\
Despite the advantages of the current PET scanners they are characterized by a number of technical and conceptual limitations. Actually, the field of view of the body that can be imaged at one shot does not typically exceed 250~mm in length~\cite{SLOMKA2016}. This means that any full-body scan  has to be merged from several subsequent, not simultaneous measurements. Therefore, the information about temporal changes in radiotracer distribution is available only for the fraction of the body within the field of view of the scanner. In the present-day PET scanners less than 1 \% of the photons emitted from a patient body are detected as a consequence of a limited axial field of view (AFOV)~\cite{CHERRY2018}. For these reasons, the concept of a total-body scanner which allows almost complete detection of the radiation emitted from the body appears naturally desirable~\cite{CHERRY2017A, CHERRY2018}. Furthermore, the total-body PET will enable decreasing in the time of diagnostics or the amount of the administered radiation dose and it may also enable more effective application of shorter living tracers. Recently different designs of total-body scanners based on the standard technologies were introduced e.g. using resistive plate chamber (RPCs)~\cite{Blanco2006}, straw tubes~\cite{LWS2005,SUN2005} and 
crystal scintillators~\cite{CHERRY2017A, CHERRY2018}. 
The total-body PET based on crystal scintillators is already in the stage of commissioning~\cite{BERG2018} and delivering first total-body images~\cite{BADAWI2019}.
\\
The J-PET (Jagiellonian-PET) project addresses the innovative application of plastic scintillators as a detection material for the PET \cite{Moskal2011A,MOSKAL2016A}. 
The application of plastic scintillators enables construction of a cost effective total-body scanner due to the less expensive detector material and the possibility of the construction of the scanner from the long axially arranged plastic strips~\cite{Moskal2011A, Moskal2014A, Moskal2015A,MOSKAL2016A,KOWALSKI2018}. Moreover the readout components are placed outside of detection chamber giving a chance for hybrid PET/MR construction. In the axial arrangement of scintillator strips, any extension of AFOV by elongation of the plastic scintillators may be achieved without significant increase of costs because the number of photomultipliers and electronic channels remains independent of the AFOV. 
Due to the low light attenuation in the plastic scintillators, the length of modules could approach even 2~m. Though, this comes with deterioration of the Coincidence Resolving Time (CRT), which decreases with elongation of modules \cite{MOSKAL2016A} it can be compensated by the registration of light escaping from the scintillators with the additional layer of wavelength shifters~\cite{SYMRSKI2017}. In addition the J-PET design enables possibility of simultaneous metabolic and morphometric imaging based on the measurement of properties of positronium atoms produced inside the body during the PET diagnosis~\cite{MOSKAL2019,NATURE2019}.
\\

 In this paper the prototype built out of 24 modules, forming a cylindrical diagnostic chamber with the inner diameter of 360~mm and the AFOV of 300~mm, is presented. The developed methods of synchronization and calibration of the entire setup composed of plastic scintillators, Photomultipliers (PMT) and Front-End Electronics (FEE) are described in detail.  The result of a simplified image reconstruction is shown in the last chapter.

\section{General concept of the J-PET scanner}
\begin{figure}[h]
      \includegraphics[scale=0.75]{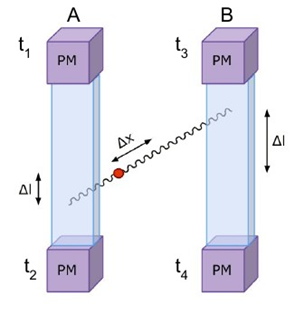}
      \label{fig:subim1}
      \includegraphics[width=4.5cm, height=3.5cm]{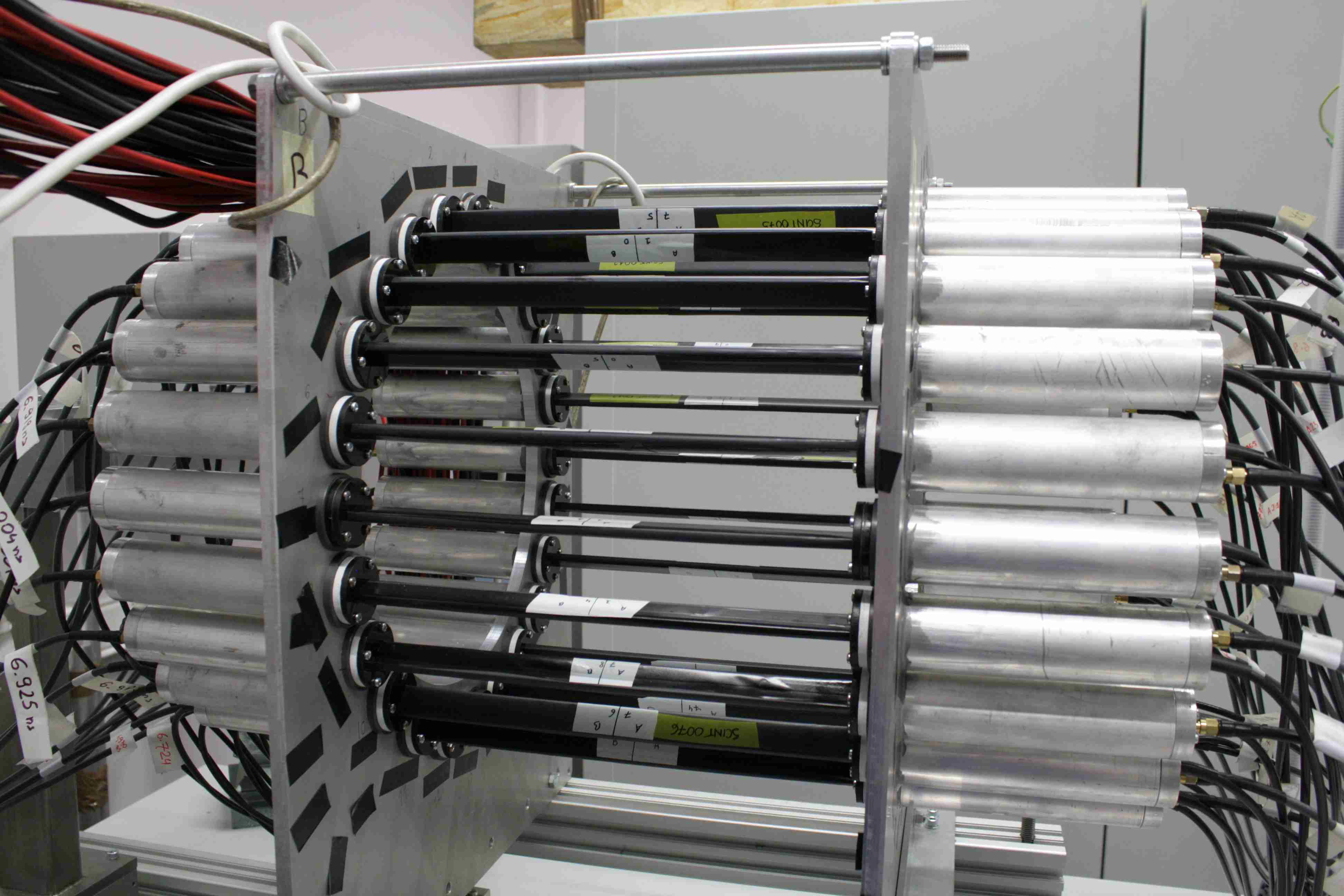}
      \label{fig:subim2}
      \caption{
(Left) Schematic representation of an annihilation point reconstruction based on measured differences between arrival times $t_{i}$ of light pulses generated in two detection modules by the annihilation gamma quanta. $\Delta$x denotes distance of the annihilation point from the middle of the LOR. (Right) 24-modules full prototype of the J-PET detector.
Scintillator strips are covered with black foil and read out by photomultipliers inserted into aluminium tubes.
}
      \label{fig:Reconstruction}
\end{figure}

J-PET exploits time information instead of energy to determine place of
annihilation. Scintillating signals from plastics are very “fast” (typically, 0.5 ns rise time and 1.8~ns decay time~\cite{saint,eljen,wieczorek2017}).
Such fast signals allow for superior time resolution and decrease pile-ups with respect to crystals detectors as e.g. LSO or BGO with decay times equal to 40 ns and 300 ns, respectively \cite{omega}.
In order to take advantage of these superior timing properties of plastic scintillators and to decrease the dead time due to the electronic signal processing in J-PET, the charge measurement corresponding to the deposited energy of the gamma was replaced with measurement of Time Over Threshold (TOT) ~\cite{Paka2017}.
The J-PET tomograph is constructed from axially arranged strips of plastic scintillators. Annihilation $\gamma$ photons with energy of 511 keV interact in plastic scintillators through the Compton effect \cite{MOSKAL2018} in which the deposited energy varies from event-to-event. Due to the low light attenuation plastic scintillators act as effective light-guides for these secondary photons produced by interaction of the annihilation radiation. Hence the examination chamber can be built out of long modules placed along the patient's body. Each plastic strip is read out by photomultipliers at two ends (see Fig. \ref{fig:Reconstruction}, left panel).
Since the readout is placed outside of the diagnostic chamber, the main cost of extending the AFOV of the scanner lays in cost of scintillating material.
The position of interaction with the photons in the scintillators can be determined from the time
difference of light signal arriving at photomultipliers placed at each end of detection module
\begin{equation}
    \Delta I =(t_1-t_2)\times \upsilon/2
\end{equation}
where $\Delta I$ denotes the distance between the interaction point and the center of the module, t$_1$ an t$_2$ stand for times of arrival of light signal at each side and $\upsilon$ is an effective velocity of light signal within the scintillator. Then, the position of annihilation along Line Of Response (LOR) can be determined using the Time of Flight (TOF) method (see Fig. \ref{fig:Reconstruction} for pictorial description)
\begin{align}
     TOF&=(t_1+t_2)/2 - (t_3+t_4)/2:  &  \Delta x&=TOF \times c/2,
\end{align}
where $\Delta x$ denotes distance of annihilation point from the middle of LOR, c stands for the speed of light, $t_1$ and $t_2$ are the times measured at the two ends of module A and $t_3$ and $t_4$ denote times registered with module B.

\section{J-PET prototype electronics, time and charge measurement}
\begin{figure}[thpb]
  \centering
  \includegraphics[width=0.50\textwidth]{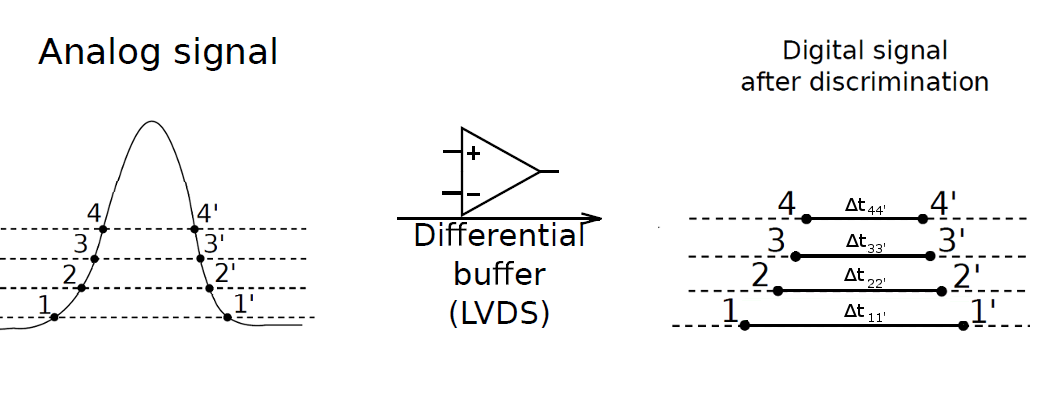}
  \caption{ Pictorial representation of electric signal probing. After signal processing, four pairs of points are acquired at four selected voltage thresholds.}
  \label{fig:signal probing}
  \end{figure}

\begin{figure}[thpb]
    \centering    
    \includegraphics[width=0.5\textwidth]{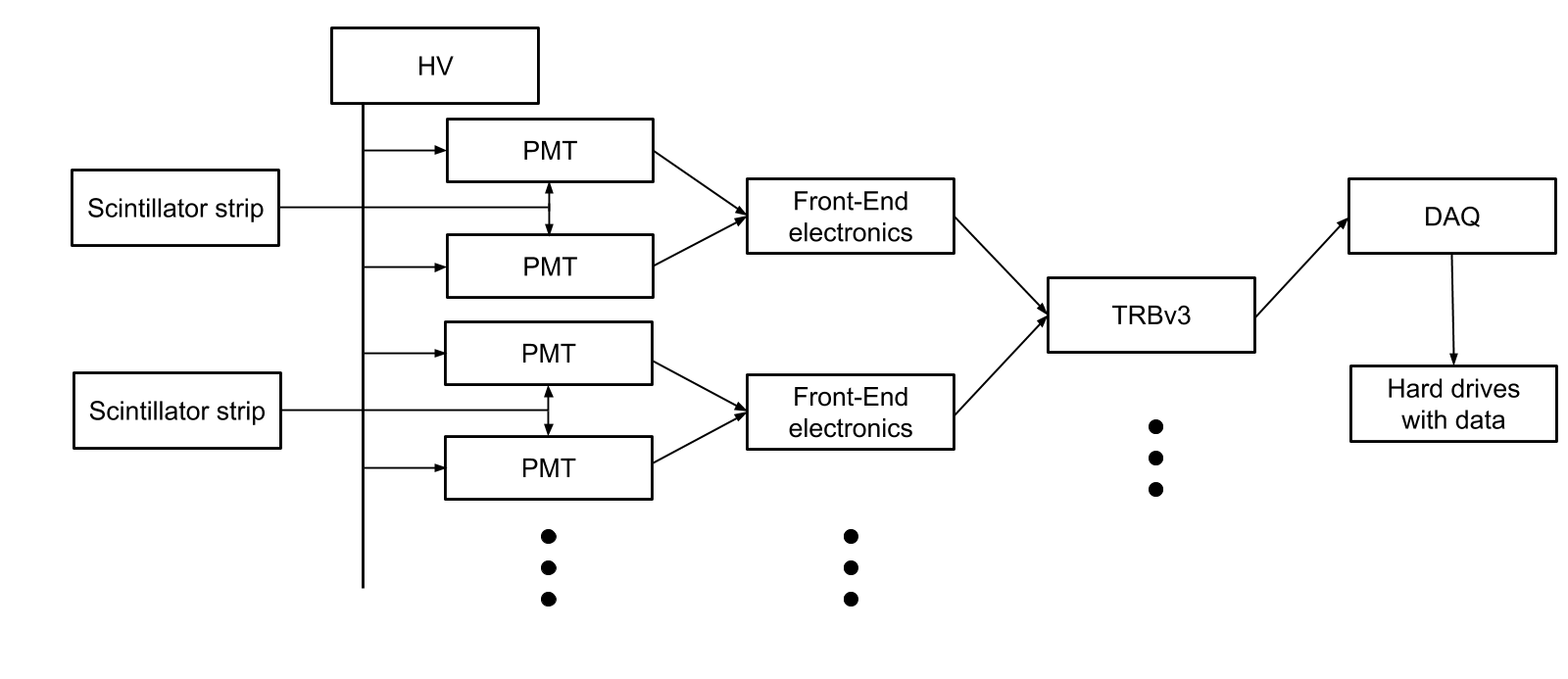}
    \caption{Block diagram of electronics supporting the 24 modular J-PET tomograph prototype. The HV denotes the high voltage supplying the photomultipliers (PMT) which are read out by the Front-End electronics connected to the Trigger and Readout Board version 3 (TRBv3) and Data Acquisition System (DAQ).}
    \label{fig:J-PET tomographs}
\end{figure}

\begin{figure}[h]
    \includegraphics[width=0.50\textwidth]{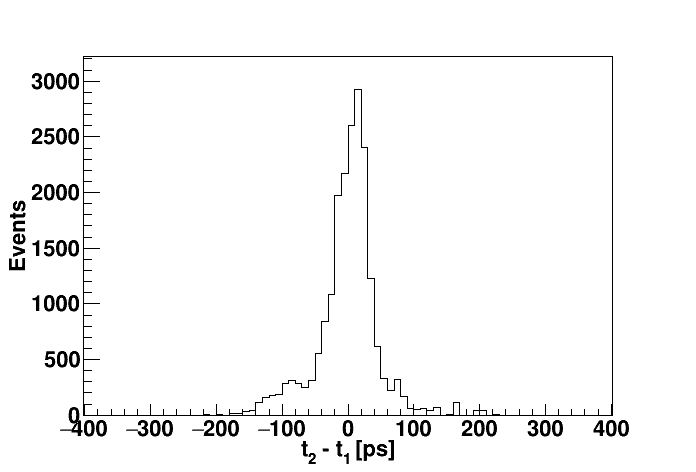}
    \caption{ 
    Time difference distribution obtained with a system equipped with FEE and TRBv3. The resolution of the obtained time difference amounts to $\sigma$ = 28.2 $\pm$ 2.5 ps.}
    \label{fig:multi-threshold board}
\end{figure}

The first operating prototype of the J-PET tomograph, shown in the right panel of Fig.~\ref{fig:Reconstruction}, consists of the 24 detection modules. A basic part of the prototype is the single detection module. It consists of 
(5 $\times$ 19 $\times$ 300~ mm$^3$) strip of BC-420 scintillator (Saint Gobain Crystals~\cite{saint})  read out by two R4998 Hamamatsu photomultipliers~\cite{HAMAMATSU} coupled optically to the scintillator with a silicone optical grease BC-630 (Saint Gobain Crystals). In order to increase the number of photons which can reach the photomultipliers, the scintillator is wrapped with the Vikuiti reflecting foil (3M Optical Systems~\cite{VIKUITI}). The lightproof of the detection module is additionally assured by a tight cover made of the Tedlar foil (DuPont~\cite{Dupont}).
\\
Electric signals from all detection modules are passively split into four, next amplified and sampled by a specially designed FEE board~\cite{Paka2017}.
It comprises 48 ABA-51563 amplifiers and 8 LTC2620 DACs to set individual thresholds fed into comparators implemented solely on a Field Programmable Gate Array (FPGA) device. The sampling of analog signals on an FPGA is executed by employing  its Low Voltage Differential Signal (LVDS) buffers as comparators~\cite{Paka2014}. It is worth to stress that also other FPGA based designs for sampling of fast signals in the voltage domain were developed recently~\cite{WON2016,WON2016B,WON2018}.
In the prototype presented in this article, the sampling of the analog voltage signal is done at four different constant thresholds at the rising and falling edges as it is depicted schematically in the left part of Fig.~\ref{fig:signal probing}. The measurement of TOT and Time to Digital Conversion (TDC) results in the digital characteristics of the probed signal, shown in the right part of Fig.~\ref{fig:signal probing}. Combination of rising and falling edges information allows for determination of signal's charge. The time determined from the crossing of the smallest threshold allows one to estimate a start time of the signal. The times measured at higher thresholds may be used to improve the precision of the start time determination e.g. by fitting a line
to the time stamps measured at the leading edge of the signal~\cite{KIM2009B,XIE2005,XIE2009A},
or by the reconstruction of the full signal waveform which may be done by fitting a curve describing the shape of the signal using either the method of library of synchronized model signals~\cite{Moskal2015A} or  e.g. the signal shape reconstruction by means of the compressive sensing theory \cite{Lech2014B, Racz2015}.
%
%
The general block diagram of electronics supporting the 24 modular J-PET tomograph prototype is shown in Fig.~\ref{fig:J-PET tomographs}. For the collective power supply of 48 photomultipliers (PMT), the CAEN SY4527 card and the CAEN SY5527 power supply were used (CAEN 2015)~\cite{Caen}.
48 analog signals from the PMTs are supplied to 4 FEE modules, which are mounted on a single Trigger Readout Board v3 (TRBv3)~\cite{Traxler2011}. The platform operates in a continuous readout mode which helps maximizing the amount of collected data without preliminary selection. The board is equipped with 5 FPGA devices (Lattice ECP3), from which one operates as a master and four others as slaves being programmed with TDC firmware. The TDCs digitize the input signals and store them in buffers, which are read out at a fixed frequency of 50 kHz. A signal to initiate the readout as well as a reference signal for precise time synchronization between the TDC is provided by the master FPGA. Collected data is then sent via Gigabit Ethernet network to the storage and further analysis~\cite{Korcyl2015, Korcyl2018a}.
\begin{figure}[thpb]
    \centering    
    \includegraphics[width=0.52\textwidth]{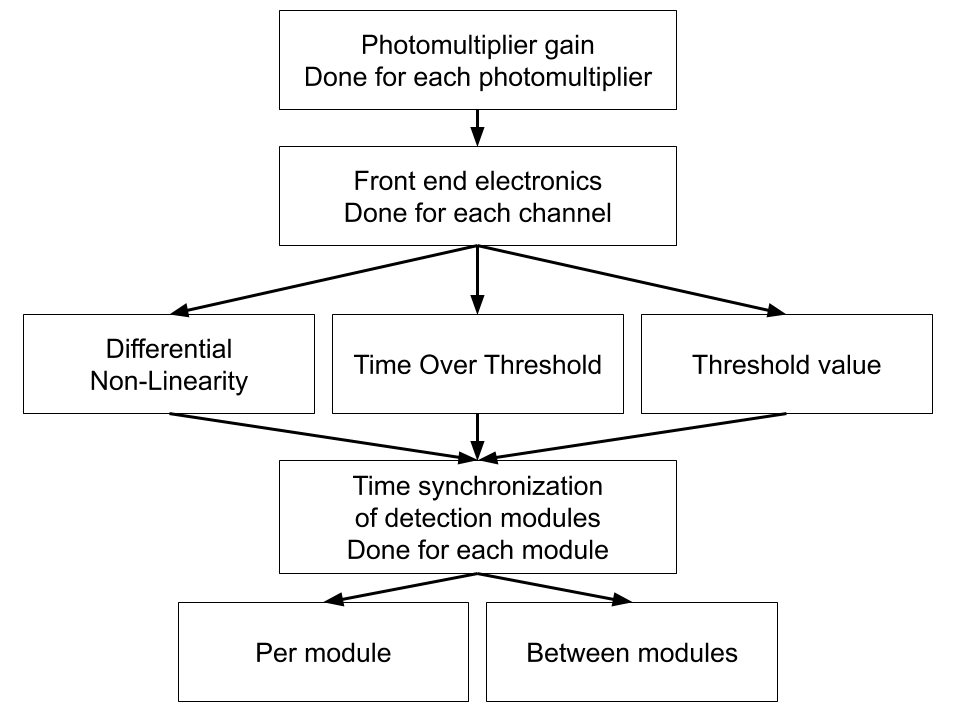}
    \caption{Block diagram of the general  flow of the J-PET detector calibration. We start with the photomultipliers gains matching, then we calibrate the differential non-linearity of the FEE, threshold values and TOT measurement. This allows for the final time calibration and synchronization of the full detector.}
    \label{fig:calib1}
\end{figure}
The described method of timing measurements by means of FPGA devices requires calibration due to following reasons. The internal carry-chain elements used as delay units have a different physical arrangement inside the integrated circuit and the pulse transition times of individual channels can vary up to a nanosecond. Also, the lengths and shapes of signal paths on printed circuits become important in the field of picosecond accuracy of time measurements. The applied calibration procedure assigns to various channels such a time shift that the time difference between them and an arbitrary chosen reference channel tends to zero. The typical time difference spread for one of the channels passing FEE and TRBv3 is shown in Fig.~\ref{fig:multi-threshold board}
giving time resolution of the order of 30~ps.
The general scheme of the calibration of the whole system is shown in Fig.~\ref{fig:calib1}, while in the next section we present calibration of the photomultipliers gains. The Front-End Electronics calibration and time synchronization of the whole J-PET detection system are discussed in Section V and VI, respectively.
\section{PMT gain calibration}
\begin{figure}[hbtp]
    \centering
    \includegraphics[width=0.37\textwidth]{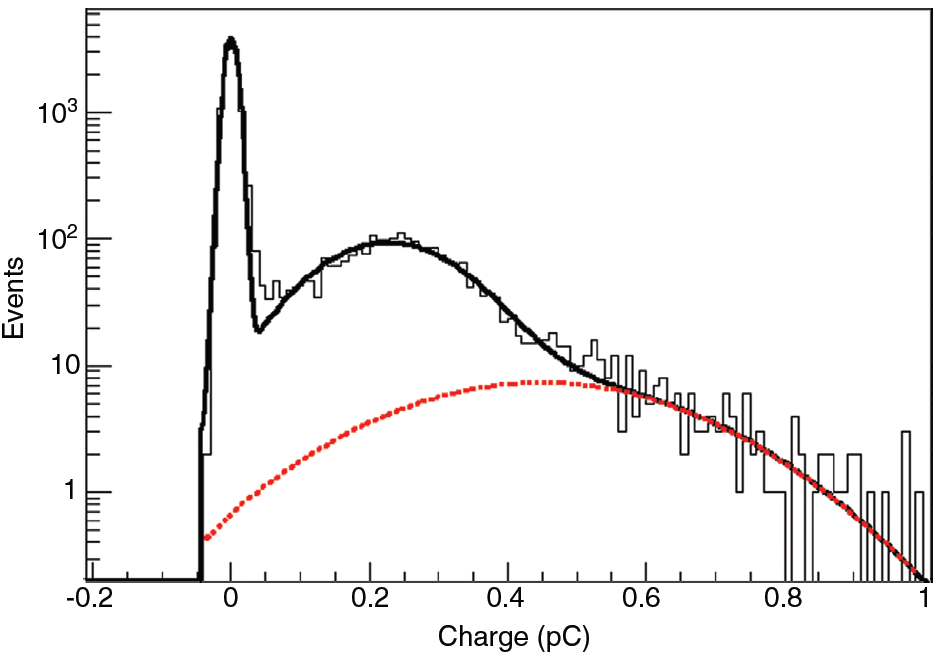}
    \includegraphics[width=0.433\textwidth]{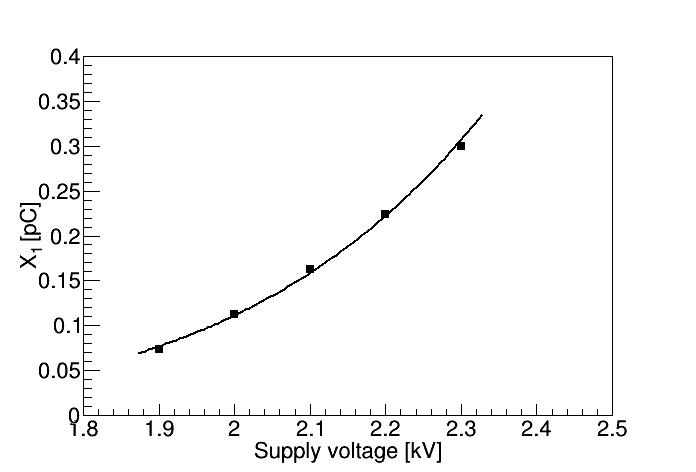}
    \caption{(Upper panel) Histogram of the charge measured during calibration of a photomultiplier with 0, 1 and 2 photoelectrons maxima (see text). (Lower panel) Example of a gain calibration curve. Average values of PMT output charges induced by one photoelectron detected as a function of the voltage applied to the photomultiplier is presented. The black continuous line denotes an exponential fit to the experimental points.}
    \label{fig:calibration}
\end{figure}

\begin{figure}[h]
    \centering    
    \includegraphics[width=0.50\textwidth]{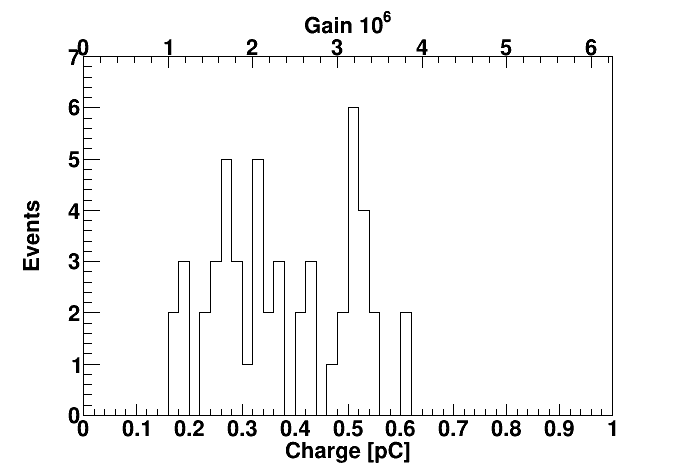}
    \caption{Values of the gain for all the R4998 Hamamatsu PMT’s obtained from gain calibration curves for the voltage of 2250~V. Note that the histogram includes results for 51 PMTs and out of them 48 were used in the prototype described in this article.}
    \label{fig: Hamamatsu1}
\end{figure}

PMT gain calibration is based on observing single photoelectrons. The method of recording individual photoelectrons has been used in many experiments 
eg.~\cite{Ronzhin2010,Baturin2006}. 
The calibrated PMT was optically coupled to a gamma irradiated scintillator. The second reference PMT, placed on the other end of the same scintillator was working in coincidence to eliminate false signals, for example, from the thermal noise of the calibrated PMT.
For a given voltage applied to the calibrated PMT waveforms of signals were collected and integrated using Riemans integral. The value of the integral was then divided by oscilloscope channels resistance used in waveforms collection, equal to 50~$\Omega$, which resulted in a charge calculation. Since the surface of the photomultiplier was obstructed in such a way that mainly one/two optical photons from the scintillator could successfully interact with its window, the charge calculated based of waveforms originates mainy from one initial photoelectron.
The histogram showing example events coming from observation of 0, 1 and 2 photoelectrons can be seen in the upper part of Fig.~\ref{fig:calibration}. Histograms of charges were fitted with a function given by Eq.~(\ref{eq3}):
\begin{eqnarray}
         F(x)&=&N_0\exp{\frac{-(x-X_0)^2}{2\sigma _0^2}}+N_1\exp{\frac{-(x-X_1)^2}{2\sigma _0^2}}\nonumber \\
    &+& N_2\exp{\frac{-(x-2X_1)^2}{2\sigma _0^2}},
    \label{eq3}
\end{eqnarray}
where N$_i$ (i = 0, 1, 2) are normalisation constants,
X$_i$ are mean (expectation) values and $\sigma_i^2$ are Gaussian variances. According to linear scaling, it was assumed that the 
expectation value for two photoelectrons (2X$_1$) is twice as large as the maximum position (X$_1$) for single photoelectron. The black curve in the upper part of Fig.~\ref{fig:calibration} is the sum of three Gaussian curves corresponding to situations when no photoelectron was registered (the first maximum),
when one photoelectron was observed (the central maximum) and when there were two photoelectrons (the submerged red line).
For charges above 0.6 pC the red line overlaps with the black one. The value of the X$_1$ parameter obtained from the fit is then used as a measure of the gain of photomultiplier.
The gain calibration curves can be obtained performing fits of
Eq.~(\ref{eq3}) to histograms measured with different voltages applied to PMT’s. The typical example
of such a curve is shown in the lower panel of Fig.\ref{fig:calibration}.
Values of gains gathered for all 48 photomultipliers used in the J-PET prototype, operated at the voltage of 2.25 kV, are shown in Fig.~\ref{fig: Hamamatsu1}. As one can see they differ by a factor of about 3.
\section{FEE calibration}
\begin{figure}[h]
    \centering    
    \includegraphics[width=0.40\textwidth]{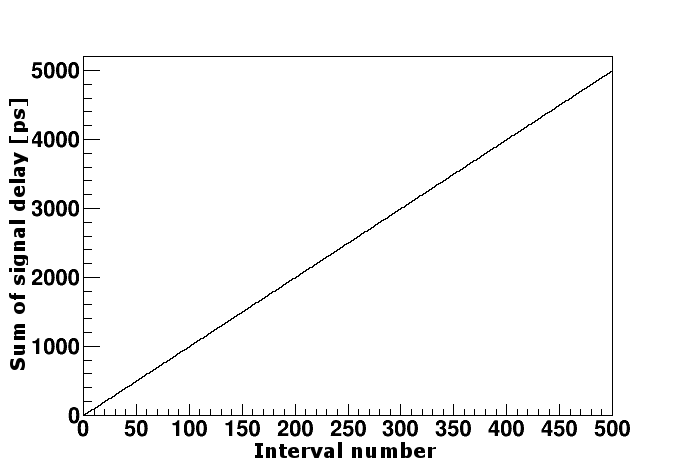}
     \includegraphics[width=0.40\textwidth]{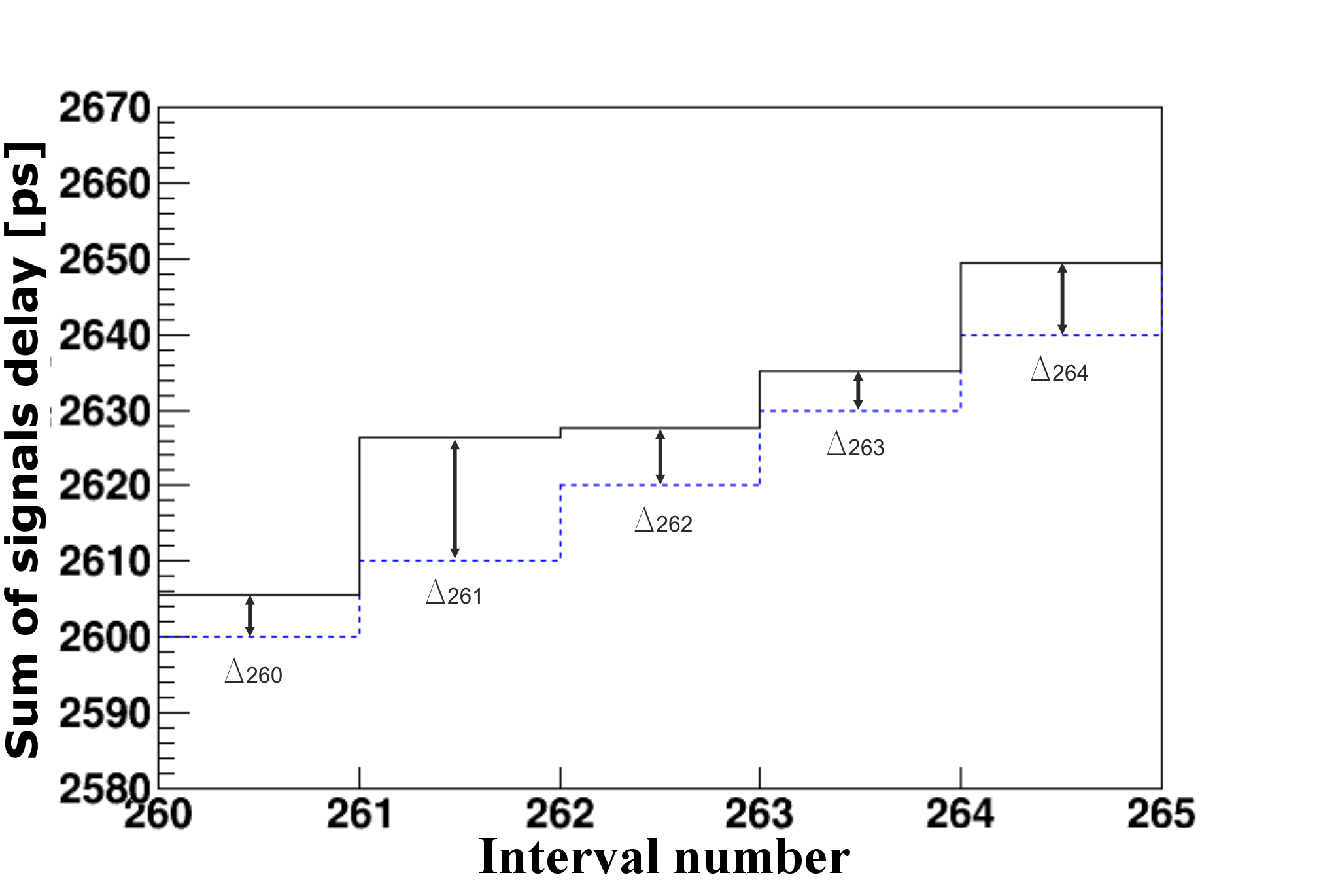}
     \vspace{0.2cm}
    \caption{TDC Calibration for Differential Non-Linearity (DNL). (Upper panel) Sum of signals delays for different interval number. (Lower panel) Zoom of the upper plot with time correction constant visualization.}
    \label{fig:DNL}
\end{figure}

As it was mentioned in Sec. 3 the J-PET Data Acquisition System (DAQ)~\cite{Korcyl2015,Korcyl2018a} is based on the TRBv3 \cite{Traxler2011} and on specially designed  FEE~\cite{Paka2014,Paka2017}.  Time measurement using FPGA is based on the signal delay resulting from its propagation through the individual elements of the chain of delays. Depending on the number of elements through which the signal has passed until the instant of measurement, we obtain different time values and calculating this time interval we assume that it is proportional to the number of passed elements. This calculation is correct as long as the propagation time of the signal through each of the chain elements is the same. In general, this assumption is not fulfilled and this problem is known as Differential Non-Linearity (DNL) of time propagation of the signal in the TDC system. The level of non-linearity for individual elements is dependent on the temperature and voltage fluctuations in the electronic components during system operation. This is one of the major factors that worsen the resolution of time measurement therefore a calibration of DNL is needed. It consists in giving to the TDC input a large number of signals which are accidental, uncorrelated with an internal clock signal and homogeneously distributed within the interval of the time measurement. From these signals a histogram of time as a function of delay chain element number is created, where time is calculated as a sum of delays at all elements counting from the beginning of the delay chain till the given element. Example of DNL correction histogram is shown in 
Fig.~\ref{fig:DNL}.

In the case of ideal elements having the same delay, the histogram should be arranged in a stepped line with the same difference in height for each subsequent interval (blue dotted line in the lower panel 
of Fig.~\ref{fig:DNL}). As can be seen in this figure, however, the differences between the intervals vary which is a sign of DNL, i.e. nonlinearity on the elements of the delays chain. Such a histogram is created for each chain of delays through which signals are passed and the information contained therein is used to correct the measured signals.

In addition to the DNL calibration it is also necessary to calibrate the measurement of the width of the signals. Information about the signal size can be obtained on the basis of time measurements. With the increasing signal charge, its width increases, and thus also the time when the signal voltage exceeds some pre-set threshold. For this measurement one needs information about two times: the time when the rising edge exceeded the level of the threshold and the time when the falling edge crosses the level of this threshold. Such measurement of the TOT allows for determination of charge with very good resolution which then can be used for rejection of noise originating from registration of  photons scattered in the body of the patient.

\begin{figure}[h]
    \centering    
    \includegraphics[width=0.40\textwidth]{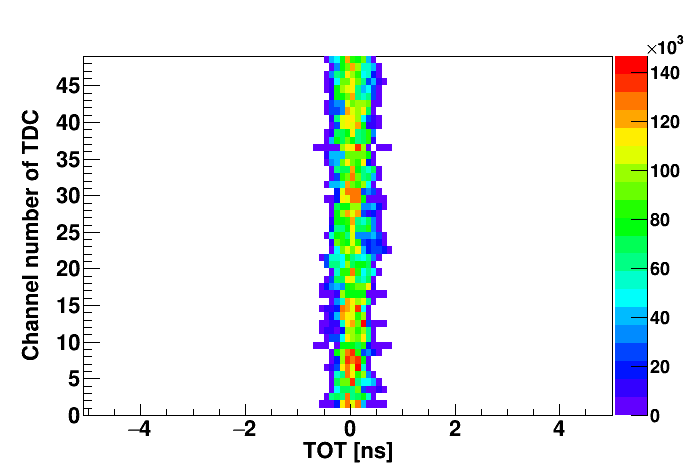}
    \caption{TOT values measured for all TDC channels of the 24-modules J-PET prototype after calibration.}
    \label{fig:TDC}
\end{figure}

Due to the very short signals from the used scintillators and photomultipliers (order of ns), delays were deliberately introduced when measuring the falling edge. In TDC, for each channel a chain of delay elements has been implemented through which the falling edge of signal must pass. This results in an artificial extension of the signal, allowing, however, the measurement of TOT for very narrow signals. The delays on different TDC channels may slightly differ from each other and this entails the need for TOT calibration. Such calibration is made using an additional oscillator placed on the TRBv3 board providing a reference signal of 10~ns wide at the input of each channel. Thanks to this, it is possible to simultaneously measure the signal width for all channels and find the value of the calibration parameters of the falling edge delays for each channel. On the basis of the mean of the measured TOT values for each channel, the values of edge times of falling signals on the given channel are corrected. After subtraction of the calibration signal width, the corrected time should give zero values on all channels. Proof of the proper operation of this procedure is presented in Fig.~\ref{fig:TDC}.

\begin{figure}[h]
         \includegraphics[width=0.40\textwidth]{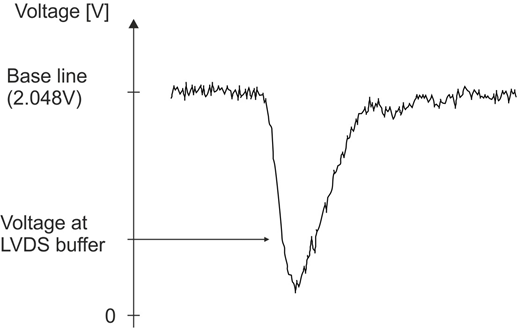}
        \includegraphics[width=0.40\textwidth]{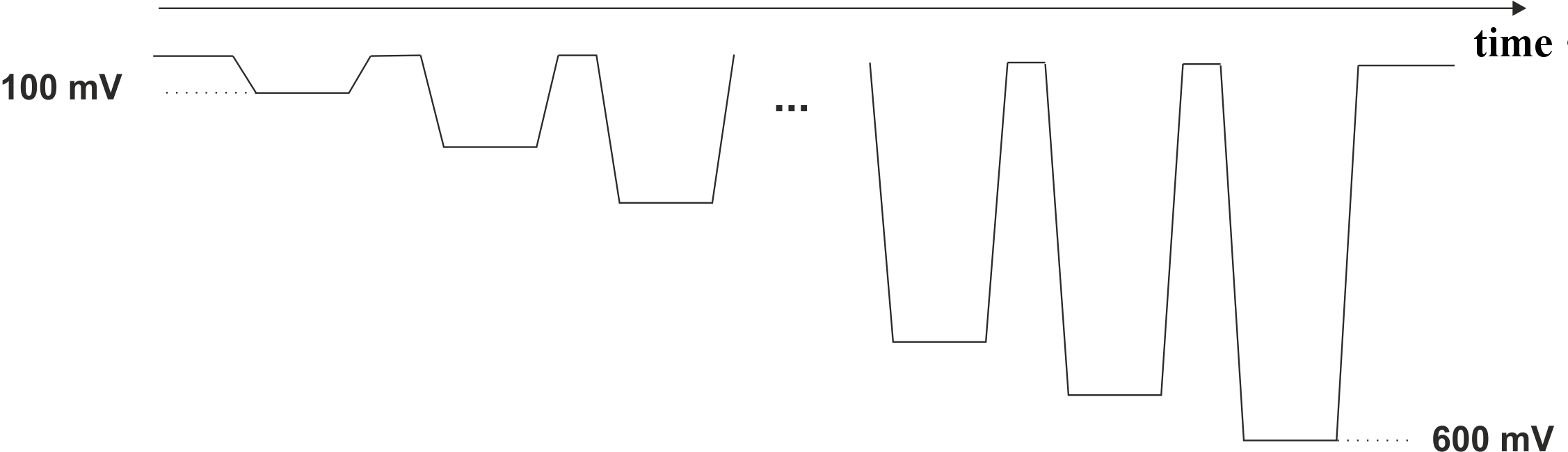}
    \caption{(Upper panel) Pictorial representation of the calibration of threshold values applied to the J-PET photomultiplier signals. The value of real threshold equals to the difference between levels of base line and voltage at LVDS buffer.; (Lower panel) Schematic view of signals sequence used for threshold level calibration.
    \label{fig:threshold level}}
\end{figure}

\begin{figure}[h]
\centering
       \includegraphics[width=0.40\textwidth]{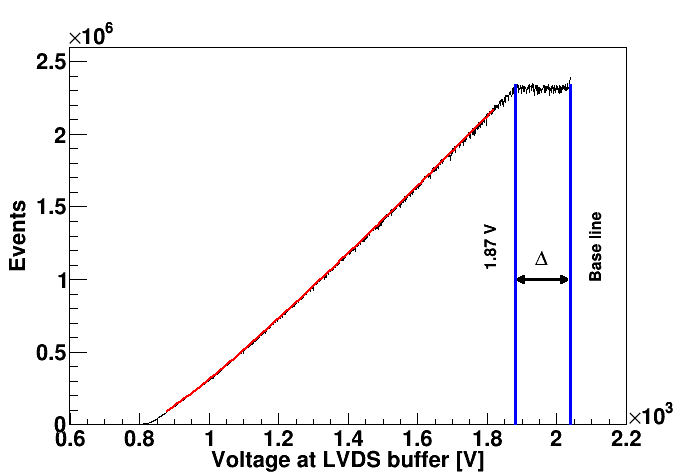}
      \includegraphics[width=0.40\textwidth]{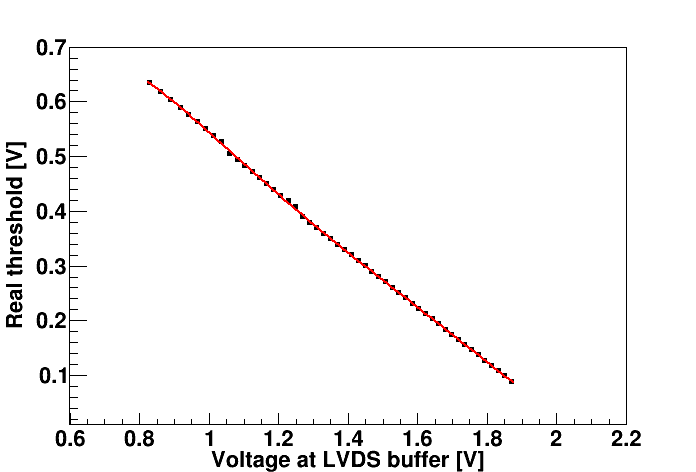}  
  \caption{(Upper panel) Number of registered events for different voltage level set at LVDS buffer with the baseline set to 2.048~V. (Lower panel) Dependence between the absolute value of the real threshold and the voltage set at the LVDS buffer. A 5th order polynomials (red solid lines) were fitted to the data for the slopes description.}
  \label{fig:registered}
\end{figure}

In the J-PET prototype dedicated FEE uses LVDS buffers to compare reference voltage (threshold level) with measured signals. The LVDS buffer work in the range from 0 V to 2 V while signals registered from photomultipliers have negative amplitudes. Therefore, to be able to apply a threshold on a signal with a negative amplitude, the base level of the signal has been shifted from zero to 2.048~V. As a result, signals of negative amplitude remain in the domain of positive voltages. The schematic representation of this procedure is presented in the upper panel of Fig.~\ref{fig:threshold level}.
\\
The relation between the actual voltage threshold sampling, the recorded pulse and
the voltage applied to the LVDS comparator input has been determined in the following way.
A sequence of pulses of variable amplitude (shown schematically in the lower part of 
Fig.~\ref{fig:threshold level}) was repeatedly sent from programmed generator to each input channel of readout electronics.
The parameters of this sequence were as follows:
\begin{itemize}
    \item amplitude: changing from -100 mV to -600 mV with a step of 10 mV
    \item rise time and fall time: 1 ns (10$\%$ \text{-} 90$\%$ of amplitude)
    \item signal width at half amplitude: 4 ns
\end{itemize}
While sending the pulse sequences from the generator to the readout electronics, the voltage
level on the LVDS buffer was changed and the number of signals that exceeded this level
was counted. An example of the dependence of the number of signals accepted as a function
of the voltage level on the LVDS buffer is shown in the upper panel of Fig.~\ref{fig:registered}. Because the
baseline has been moved up to 2.048~V, for negative signals, a smaller threshold value means
the threshold applied at a higher signal amplitude, according to the upper part of 
Fig.~\ref{fig:threshold level}. In order to describe dependence of the number of counts on the voltage on the buffer, a 5${^{th}}$
degree polynomial was fitted to the data. The flat part of the chart above 1.87~V level corresponds to the threshold values for which all pulses from a single sequence with the amplitude greater than 100~mV were registered.
\\
By combining the information on the number of sequences, the number of pulses in the sequence and how many pulses from the sequence were registered on the threshold with a given voltage value, it is possible to convert the voltage level set on the LVDS buffer to the actual threshold applied to the signal. An example of determination of the absolute value of the real threshold (a distance from the baseline at 0~V) as a function of voltage at the LVDS buffer is shown in the lower panel of Fig.~\ref{fig:registered}.
\section{Synchronization of the J-PET prototype}
\begin{figure}[h]
\centering
       \includegraphics[width=0.40\textwidth]{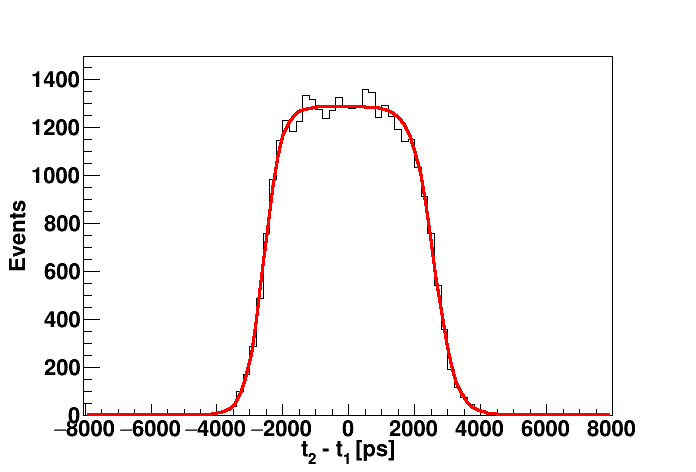} 
         \includegraphics[width=0.40\textwidth]{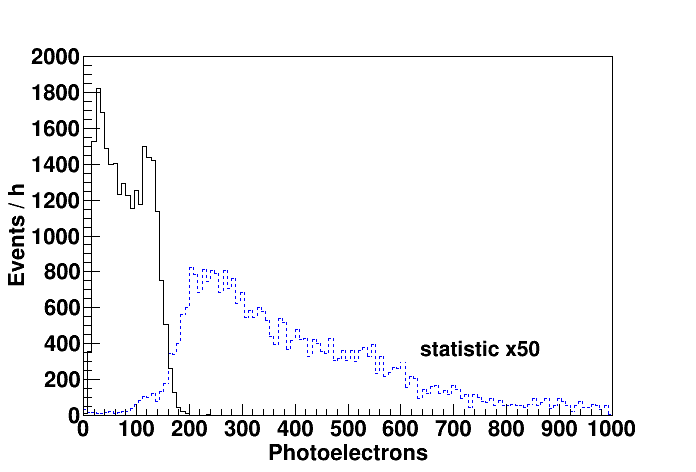}
    \caption{(Upper panel) Example of distribution of the time difference between signals arrival to the two
photomultipliers located at the ends of a scintillator irradiated with cosmic rays.
The red line shows a fit of the double Fermi function. (Lower panel) Distribution of the number of photoelectrons per event observed using a $^{22}Na$ radioactive source with activity of 17.3~MBq (black solid histogram) and cosmic radiation (blue dashed histogram) registered simultaneously with the annihilation gamma quanta scaled up by a factor of 50 for better visibility.}
    \label{fig:distribution}
\end{figure}

\begin{figure}[h]
      \includegraphics[width=0.34\textwidth]{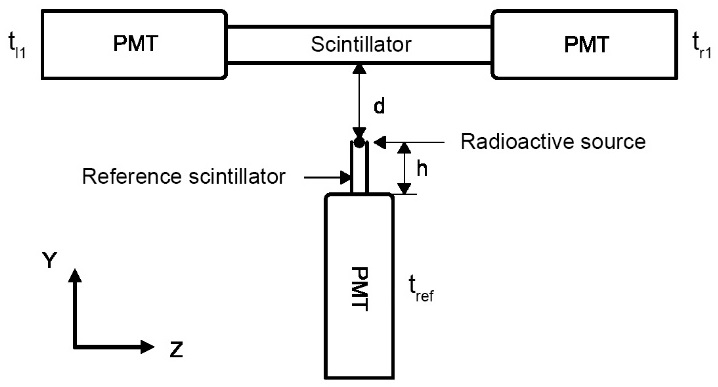}  
      \includegraphics[width=0.2\textwidth, angle=90]{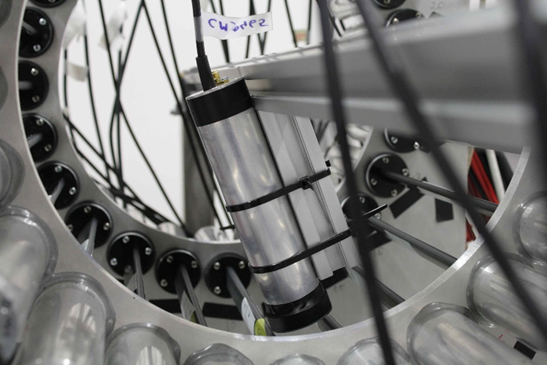}   
    \caption{ (Left) Schematic view of the method of the J-PET prototype synchronization using the reference detector. (Right) Reference detector mounted at the rotating arm inside the prototype barrel.}
    \label{fig:refrence detector}
\end{figure}

Adjustment of relative time between all elements of the detection system is necessary
in order to be able to reconstruct the place of interaction of the gamma photon with the scintillator, as well as the location of positron-electron annihilation. The time synchronization of the J-PET tomograph prototype has been divided into two stages (see Fig.~\ref{fig:calib1}):
\begin{itemize}
    \item time tuning of a single detection module,
    \item mutual time coordination of all detection modules.
\end{itemize}
The first stage of calibration of two photomultipliers in a single detection module is
based on the use of cosmic rays, which uniformly irradiate the scintillator over its entire
length~\cite{BAMScosm}. Time of signal of left and right photomultipliers can be expressed  as:
\begin{eqnarray}
         t_l = t_{hit} +\frac{z}{\upsilon}+t_{off_{l}}
         \label{eq4}
         \\
         t_r = t_{hit}+ \frac{L - z}{\upsilon}+t_{off_{r}},
        \label{eq5}
\end{eqnarray}
where $t_{hit}$ is the time in which the interaction with the scintillator occurred, $L$ is the length of the
scintillator and $z$ is the position of the gamma quantum interaction along the scintillator.
The t$_{off_l}$ and t$_{off_r}$ times are fixed time offsets for the left and right photomultiplier, respectively, resulting from propagation of the signal by FEE and cables. The speed of light $\upsilon$ in the scintillator was determined using an independent method described in~\cite{Moskal2014A}. 
When calculating the difference between the two times defined in Eqs.~(\ref{eq4}) and~(\ref{eq5}):
\begin{eqnarray}
         t_r - t_l = \frac{L}{\upsilon} - \frac{2z}{\upsilon} + t_{off_{r}} - t_{off_{l}} 
\end{eqnarray}
\\
we get relative values of time constants for photomultipliers from given detection module (t$_{off_p}$ - t$_{off_l}$ ). Thus, by determining the value of the shift from the time difference spectrum, we can obtain the correction values of the time constants by correctly locating the central distribution position. The mean value of counts or the median in the distribution are not
good estimates because the photomultiplier efficiency at either ends of the scintillator may
differ, biasing the time calibration. Therefore, in order to determine the center
position of the two edges of the time difference spectra, two functions (known as logistic function, sigmoid or Fermi function) were adjusted
with the following formulas:
\begin{eqnarray}
         f_F(x) = \frac{P_0}{\exp{\frac{x - P_1}{P_2}}+1}+P_3
\end{eqnarray}
\\
where the parameters P$_0$, P$_1$, P$_2$, P$_3$ correspond respectively to the maximum value of the function, the center of the edge, the edge inclination, and the minimum value of the function. On the basis of the position values of the centers of the two edges, it is possible to calculate how much the spectrum should be moved so that it is symmetrical with respect to the zero value. An example of the distribution of the difference in the time of registration of particles from cosmic radiation using two photomultipliers, after correction, is shown in the upper part of Fig.~\ref{fig:distribution}.
\\
An advantage of this method of synchronization is that it can be done simultaneously with a patient scan because energy of gamma photons from positron annihilation is different from the cosmic ray energies such that both components can be separated as shown in the lower part of Fig.~\ref{fig:distribution}.
\\
The concept of the time synchronization method of modules is based on the principle of transitivity, in our case on comparison with the reference point. Time synchronization between detection modules was made using a rotating sodium radioactive source together with a reference detector. The reference detector was a narrow and elongated (5 $\times$ 5 $\times$ 19~ mm$^3$) BC-420 scintillator optically coupled with an additional photomultiplier. The geometry of the reference scintillator forced the self-collimation of photons from the $^{22}Na$ source, preferring photons moving close to the longitudinal axis of the scintillator to reach the reference photomultiplier. 
The system used for mutual synchronization of detection modules is schematically presented in the left part of Fig.~\ref{fig:refrence detector}.

The right part of Fig.~\ref{fig:refrence detector} shows the reference detector mounted on the rotating arm inside the J-PET tomograph prototype. The rotation allows to achieve the configuration shown in the left part of Fig.~\ref{fig:refrence detector} with respect to each of 24 detection modules.
\\
Time of signals from left and right photomultiplier can be written analogously to formulas (4) and (5). However, in this case, the reference time $t_{hit}$ is the time of emission of photons from the source located on the reference detector. Therefore, it is necessary to add the time needed to travel the path $d$ from the source to the detector module scintillator. Then the measured time consists of the following elements:
\begin{eqnarray}
         t_{l1}= t_{hit}+\frac{d}{c}+\frac{z}{\upsilon}+t_{off_{l1}},
         \\
         t_{r1}= t_{hit}+\frac{d}{c}+\frac{L - z}{\upsilon}+t_{off_{r1}},
\end{eqnarray}
\\
The time for the reference detector can be written as:
\begin{eqnarray}
         t_{ref}=t_{hit}+\frac{h}{2\upsilon}+t_{off_{ref}},
\end{eqnarray}
\\
where t$_{off_{ref}}$ are fixed time values resulting from signal propagation in FEE and cables. The factor $\frac{h}{2\upsilon}$ describes the average time it took for the light to travel inside the reference scintillator before reaching the photomultiplier. In general, the gamma photon can react in various places along the scintillator. This results in the  variation of the obtained time, but does not alter the average value which was taken as a constant for all performed measurements. Taking into account the synchronization of a single module described previously, knowing the constant values of times t$_{off_{l1}}$ and t$_{off_{r1}}$ , the propagation time of the signal through FEE and cables for a single module in its entirety is following:
\begin{eqnarray}
         \frac{t_{off_{r1}}+t_{off_{l1}}}{2}=t_{off_1}
\end{eqnarray}

\begin{figure}[h]
\centering
\includegraphics[width=0.43\textwidth]{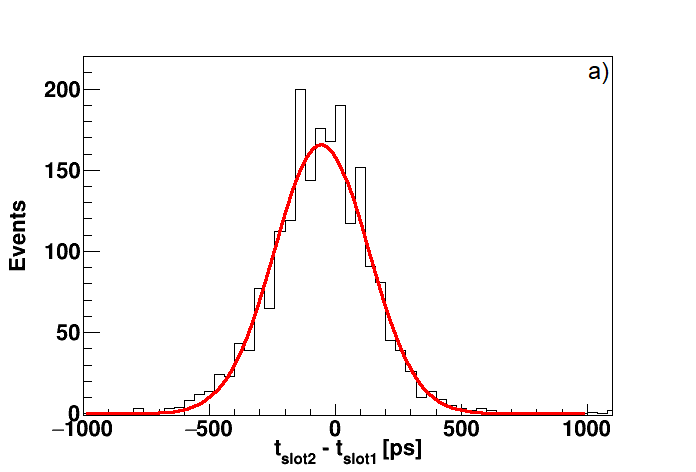}
   \includegraphics[width=0.43\textwidth]{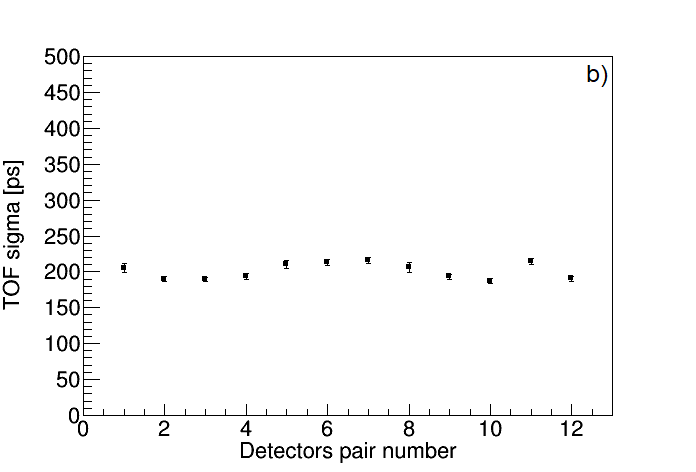}
\includegraphics[width=0.4\textwidth]{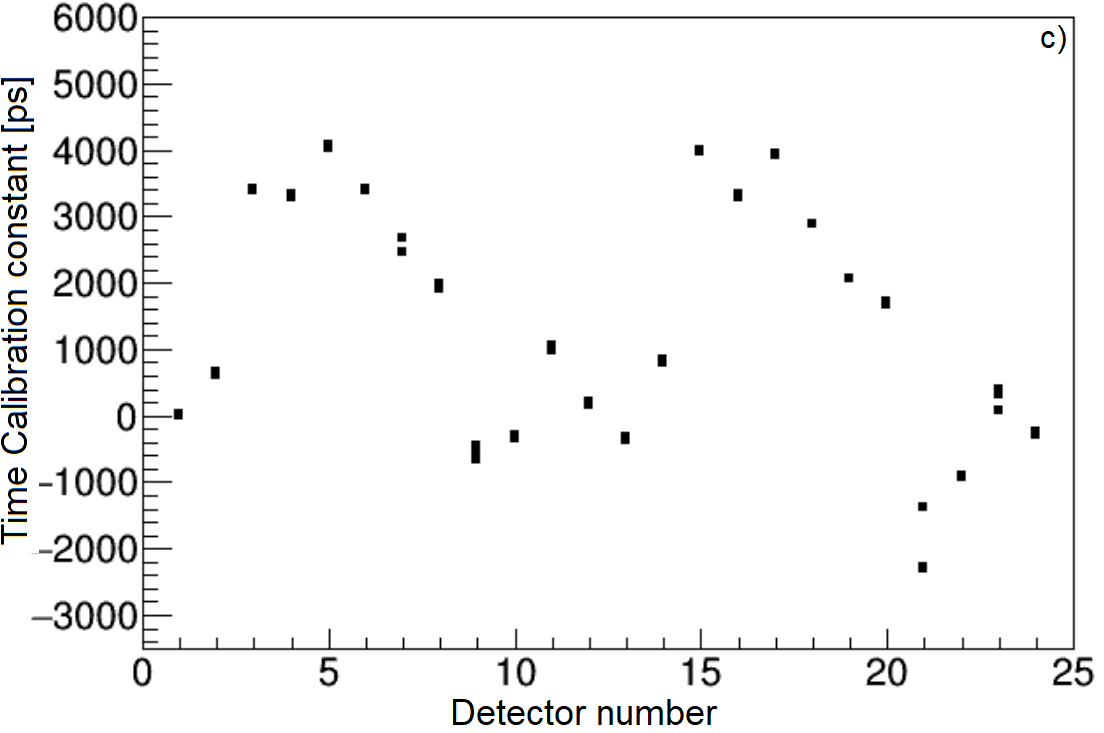}
    \caption{a) Coincidence time resolution for an exemplary pair of modules. Superimposed red line indicates a fitted Gaussian function with $\sigma$ = 187~ps. b) Coincidence time resolution for all facing pairs of modules. c) Values of the calibration constants obtained after synchronization of the whole detector with respect to the first detection module for all the four thresholds used.}
    \label{fig:gaussian}
\end{figure}

Going into a generalization for the entire prototype of the J-PET tomograph, we label detector modules with indices $i$ and $j$. Then subtracting time from the reference detector and
time from a given detection $i^{th}$ module we obtain:
\begin{eqnarray}
         t'_i=t_{ref}-\frac{t_{ri}+t_{li}}{2}=\frac{h}{2\upsilon}-\frac{d}{c}-\frac{L}{2\upsilon}+t_{off_{ref}} - t_{off_i}.
\end{eqnarray}
\\
As a result, it is possible to determine the relation between the measurement time from
any two detection modules through a reference detector, and thus to determine the relative
times and synchronization of these modules. Generally for modules $i$, $j$ it can be written as:
\begin{eqnarray}
         (t_{ref}-\frac{t_{ri}+t_{li}}{2})-(t_{ref}-\frac{t_{rj}+t_{lj}}{2})=~t_{off_j} - t_{off_i}.
\end{eqnarray}
\\
The presented method of time calibration allows to synchronize each module with respect to one arbitrarily selected detector. Values of the calibration constants obtained after synchronization of the whole J-PET prototype with respect to the first detection module are presented in Fig.~\ref{fig:gaussian}~c). This allows, of course indirectly, to reconstruct the place of annihilation of a positron with an electron in the internal space of the tomograph, and thus also in the patient's body. 
The parameter characterizing the precision of the tomographic image obtained on the basis of time information is Coincidence Resolving Time (CRT). It was determined by measuring the time difference of registration of annihilation gamma photons by pairs of modules located in the prototype barrel directly opposite to each other. In Fig. \ref{fig:gaussian}~a) an example of the measured time difference distribution and fitted Gaussian function with $\sigma$~=~187~ps is presented. 
The obtained value corresponds to CRT~=~439~ps (equivalent to FWHM). The obtained value of $\sigma$ can be translated to the position reconstruction accuracy $\sigma(\Delta l) =$~18~mm which is fairly independent of the reconstructed position.
%
\\
During the measurement a collimated ${^{22}}$Na source was placed in the geometric center of the J-PET prototype, the threshold level of the trigger was set to -200~mV. Fig.~\ref{fig:gaussian}~b) shows results of the measurement of coincidence time resolution for all twelve facing pairs of modules. As a result of the Gaussian function fit to distributions of time difference for all pairs of modules, the average coincidence time resolution $\sigma=$~208~$\pm$~4~ps 
(CRT~=~490~$\pm$~9~ps) was obtained, which is comparable to the best currently available scanners~\cite{SLOMKA2016,Van2016}.
\\
\section{Event selection and simplified image reconstruction}

For the analysis of acquisited signals a dedicated analysis framework was developed~\cite{Krzemien2015,KRZEMIEN2016}. The registered signals may originate from a single electron-positron annihilation events or from hits caused by gamma photons from different annihilations. This either true or accidental coincidences may be additionally influenced by scattering of photons prior to the registration in other detectors or in the radioactive source material. On the basis of an extensive modelling simulations~\cite{Kowalski2015,KOWALSKI2018} 
we have selected events for further analysis using several conditions. First of all we consider events when both photomultipliers in a single detection module provide a signal. Next both of such events should occur in two detection modules above some adjusted energy threshold which was optimized for the ratio between the number of true and scattered coincidences. The absolute value of the TOF from detection modules in coincidence was required to be less than 3~ns (time of flight of the gamma photon along the diameter of the scanner) since radioactive source was in the centre of the J-PET prototype.

\begin{figure}
    \centering
     \includegraphics[width=0.5\textwidth]{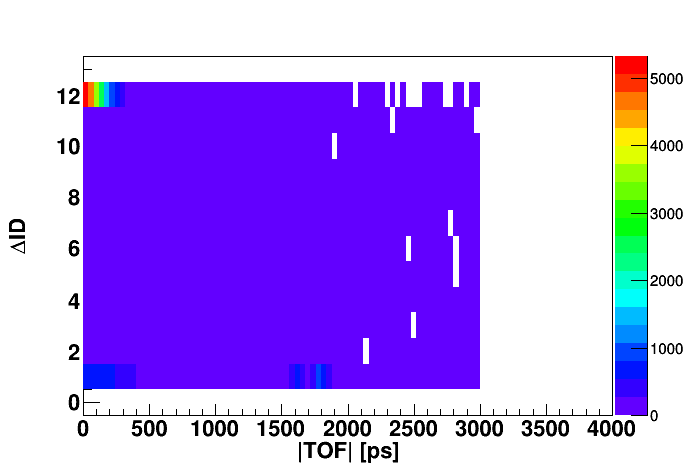}   
        \includegraphics[width=0.35\textwidth]{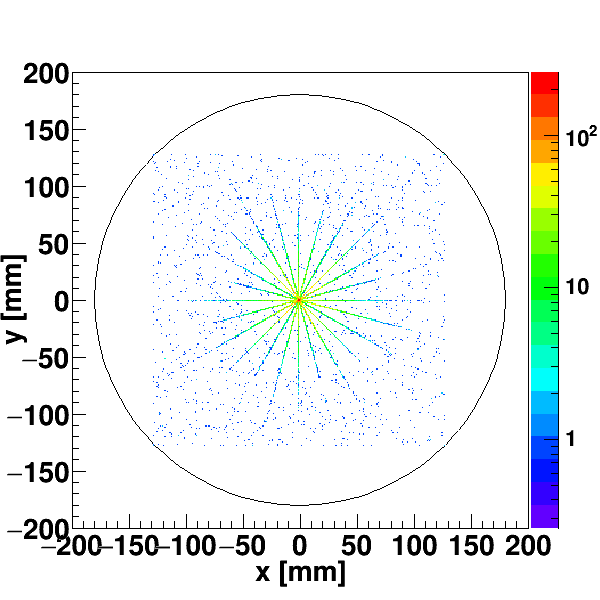}    
    \caption{(Upper panel) Correlation of the modules ID difference ($\Delta ID$) and Time of Flight (TOF) of gamma quanta measured between two detection modules for all types of coincidences, including accidental and scattered ones. (Lower panel) Result of the simplified point-like source image reconstruction in X-Y plane of the J-PET prototype. Size of the prototype is shown by the circle.}
    \label{fig:prototypr}
\end{figure}

In order to improve selection criteria which would allow a suppression of the detector-scattered and source-scattered coincidences, we have performed modelling studies of the correlation between the detector’s identity numbers ID and the time differences between the registered signals TOF \cite{Kowalski2016}. The scintillator identifiers ID increase monotonically clockwise in the range from 1 to 24 and the differences of detection module
numbers in coincidence $\Delta ID$ were calculated as follows: 
$\Delta ID = min(|ID_1 - ID_2|, 24 - |ID_1 - ID_2|)$,
where ID$_1$ and ID$_2$ denote ID of scintillator modules. For all coincidences, 2-dimensional histogram of registration time differences between subsequent scatterings TOF and scintillator identifiers differences $\Delta ID$ were calculated. This histogram is presented in the upper
part of Fig.~\ref{fig:prototypr}. The Maximum number of $\Delta ID$ is 12 which is the case when detection module lie exactly on the opposite sides relative to the detector centre. True coincidences are located in the region of low TOF and high $\Delta{ID}$. 
\\
One of the methods of reconstruction of tomographic images based on the collected data is to perform a simplified reconstruction. This can be accomplished using the idea of determining
the place of positron-electron annihilation described in Sec. 2. In this method, it is assumed that the Line Of Response (LOR) passes through the geometrical centers of cross-sections of both scintillators. No additional reconstruction data algorithms are used to build
tomographic images. It is only a collection of annihilation points, which are obtained on the
basis of the measured values of signal arrival times. In the lower part of Fig.~\ref{fig:prototypr} the
reconstructed image of a single point-like source located in the centre of the prototype detector
is shown.
\section{Conclusion}
In this paper we have described the first operating prototype of the J-PET tomography scanner built from 24 plastic scintillator strips. The 300~mm long scintillator strips were arranged in a barrel shape with 360~mm~diameter. The signals from each of the strips were recorded by two  photomultipliers  coupled optically with the scintillator material at the opposite ends of the strips. All signals were probed at four voltage levels by front-end boards and processed by dedicated Trigger and Readout Board providing time and TOT measurements.  The prototype was built in order to progress from basic single module to a system where one has to control many units,  to test electronic readout of the whole system and to develop calibration and synchronisation procedures. We found that the coincidence resolving time (CRT) for this prototype is equal to 490 $\pm$~9~ps which is comparable to the best commercial scanners. Taking only 24 detection module units turned out to be enough to reconstruct a  point-like positron source. These studies demonstrate that a full scale prototype aiming for a whole human body scan is in reach.\\
Recently the first total-body PET based on crystal scintillators was taken into operation in Sacramento~\cite{BADAWI2019}.  However, the high costs limits its dissemination not only to hospital facilities but even to medical research clinics~\cite{Majewski2020,Vanderberge2020}. In this article we presented a prototype of the cost-effective method to build a total body PET based on plastic scintillators. Prospects and clinical perspectives of total-body PET imagining using plastic scintillators are described in Ref.~\cite{Stepien2020}.
\section*{Acknowledgement}
The authors acknowledge technical and administrative support of A. Heczko, M. Kajetanowicz and W. Migda{\l}. This work was supported by The Polish National Center for Research and Development through grant INNOTECH-K1/IN1/64/159174/NCBR/12, the Foundation for Polish Science through the MPD and TEAM POIR.04.04.00-00-4204/17 programmes, the National Science Centre of Poland through grants no. 2016/21/B/ST2/01222, 2017/25/N/NZ1/00861, the
Ministry for Science and Higher Education through grants no. 6673/IA/SP/2016, 7150/E338/SPUB/2017/1 and 7150/E-338/M/2017, and the Austrian Science Fund FWF-P26783.
\bibliographystyle{IEEEtran}
\bibliography{IEEE_ref2}

\end{document}